# Plane polarized-longitudinal Phonons in realistic low dimensional systems

V.K. Jindal

*Department of Physics, Panjab University, Chandigarh-160014, India*
*e-mail address: jindal@pu.ac.in*

A realistic one-dimensional system has not only longitudinal phonons, but also possible transverse modes, which derive their restoring force from longitudinal interaction. We show that transverse motion results in a quartic displacement term in transverse direction as the first non-vanishing term in the potential energy. This results in solution of a composite longitudinal motion superimposed by a transverse motion propagating along the length direction identified as a plane polarized phonons. Interestingly, solutions of the quartic nonlinear equation have been expressed accurately, though approximately in terms of sinusoidal solutions by modifying the periodicity of sin function. The phonons along the transverse direction, now with a weakened frequency compared to the longitudinal has interesting impact- it gives rise to negative Gruneisen parameter with a value of -1 and is responsible for negative thermal expansion in the low temperature regime. Similar results of graphene sheet based on consideration of transverse (surface ripple like) modes to the planar direction, provides explanation to the observed negative thermal expansion in low temperature regime. The concept of plane polarized phonons seems new and interesting. All dynamics of atomic motion, despite involving quartc nonlinear equation is expressible in terms of simple harmonic motion. The most important feature of the transverse modes of an open surface or chain is their dependence on lateral or longitudinal modes, and as soon as more chains or more surfaces are added, bulk interactions are initiated and longitudinal dependence of transverse motion is lost and so also very distinguishing thermodynamic properties.

Phonons in one dimensional linear chain of atoms is one of the simplest problems and has been solved in the nearest neighbor and harmonic approximation in practically all elementary books on solid state physics. Generally one refers to one dimensional crystals for the sake of simplicity and as a guide to understanding complicated three dimensional and complex crystalline materials. However due to recent practical examples of one dimensional and two dimensional materials synthesized in such forms, there has been a tendency to apply the simplified results to these materials. Naturally, only *abinitio* results often win over the quantitative results and the conceptual part gets lost. The aim of this letter is focused on realizing what is grossly missing in applying assumed oversimplified models and to address the origin of newer effects as a result of consideration of newer modes. In the following, we focus only on a one dimensional crystal before making some remarks about two-dimensional crystals.

Consider a one dimensional crystal of monoatomic atoms, with each of mass m and lattice parameter, a as shown in Fig.1.

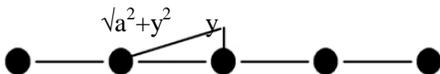

a  $(u_{s-1},y_{s-1})$  $(u_s,y_s)$   $(u_{s+1},y_{s+1})$

Fig.1. One dimensional chain of $N$ atoms with $a$ as lattice parameter and displacement $(u_s, y_s)$. $u$ shown as along the chain and $v$ along one transverse direction, $y$.

Assuming that the atoms get displaced from their equilibrium positions along length direction i.e $x$ axis, as well as along the two transverse directions, we write the following expression for interatomic potential energy in terms of displacements assumed small, retaining first non vanishing terms

$$\varphi(a+u,y,z) = \varphi(a) + \frac{1}{2}u^2\frac{\partial^2\varphi}{\partial a^2} + \frac{1}{8}\frac{y^4}{a^2}\frac{\partial^2\varphi}{\partial a^2} + \frac{1}{8}\frac{z^4}{a^2}\frac{\partial^2\varphi}{\partial a^2} \qquad (1)$$

We could have retained cubic anharmonic term in Eq. (1) involving $u^3$, that will provide coupling between transverse and longitudinal motion, but that is a lower order effect and is being ignored for the present discussion. The two additional terms on the RHS account for two transverse modes and are responsible for new additional quartic terms. We concentrate on one transverse mode and re-write as

$$\varphi(\sqrt{a^2+y^2}) = \varphi(a) + \frac{1}{8}\frac{y^4}{a^2}K_l \qquad (2)$$

where $K_l = \frac{\partial^2 \varphi}{\partial a^2}$, and is the force constant in the longitudinal direction, and results in the following equation of motion:

$$m\frac{\partial^2 y}{\partial t^2} = -\frac{1}{2}\frac{y^3}{a^2}K_l \qquad (3)$$

The solutions to this equation are obtained in the form,

$$\frac{\partial y}{\partial t} = v = \pm\sqrt{\frac{K_l}{4a^2 m}}\ y_0^2\left[1 - \frac{y^4}{y_0^4}\right]^{1/2} \qquad (4)$$

where $y_0$ defines the displacement when $v = 0$. The solution of $y(t)$ as a function of time is given by rewriting Eq.(4) as

$$\int \frac{dy}{\left[1 - \frac{y^4}{y_0^4}\right]^{1/2}} = \pm\frac{\omega_l^0}{2a}y_0^2 t,$$

Or in dimensionless variables,

$$\int \frac{dY}{[1 - Y^4]^{1/2}} = \pm\frac{\omega_l^0}{2a}y_0 t = \pm\tau \qquad (5)$$

where $\omega_l^0$ is the longitudinal frequency of the harmonic oscillator, $\omega_l^0 = \sqrt{\frac{K_l}{m}}$ and $\tau = \frac{\omega_l^0}{2a}y_0 t$.

The solution to Eq. (5) has been found to be expressed in terms of Jacobi elliptical functions [1] denoted by $sn(\tau, m)$ where m is a parameter equaling -1 in this case and written as

$$Y(\tau) = sn(\tau, -1)$$

These Jacobi elliptic functions for $m=0$, $sn(\tau, 0)$ are equal to circular functions $\sin(\tau)$.

The solutions are periodic and are represented in Fig.2. Since the functions $sn(\tau, -1)$ are very much identical to $\sin(\tau)$ (compare Fig. 2a and 2b), with one major difference in periodicity, we tried to scale the period of $sin$ function by a factor $J = 1.19814$, the two become practically equal as on graphical scale both just overlap (Fig. 2c). We thus have been able to establish that by rescaling the $sin$ curve periodicity, the $sn$ and $sin$ functions give highly identical results at least on graphical scale.

Therefore, the problem of transverse motions has been reduced to sinusoidal solutions like that of a harmonic oscillator. The elliptical functions for tranverse displacement encountered in the solution of quartic potential terms are seen to be numerically close to harmonic solutions but with a modified period of sin curve reduced from $2\pi$ to $2\pi/J$

It may be remarked that the equivalence of elliptic solutions to a sine curve is only approximate, though very close as the two graphs overlap. The resulting transverse frequency is given by

$$\omega_T^0 = J\frac{\omega_l^0}{2a}y_0, \qquad (6)$$

which is amplitude dependent.

Therefore it is proposed that the displacement in transverse motion can be considered to be governed by sinusoidal displacement alongwith the longitudinal motion. Writing these in nearest neighbor approximation,

$$m\frac{\partial^2 u_s}{\partial t^2} = -K_l[2u_s - u_{s+1} - u_{s-1}] \qquad (7a)$$
$$m\frac{\partial^2 y_s}{\partial t^2} = -K_T[2y_s - y_{s+1} - y_{s-1}] \qquad (7b)$$
and
$$m\frac{\partial^2 z_s}{\partial t^2} = -K_T[2z_s - z_{s+1} - z_{s-1}] \qquad (7c)$$

where $K_l = \frac{\partial^2 \varphi}{\partial a^2} = m\omega_l^{0\,2}$ and

$$K_T = m\omega_T^{0\,2} = mJ^2\left(\frac{\omega_l^0}{2a}y_0\right)^2 \qquad (8)$$

and assuming, $u_s = u_0 e^{-i(\omega t + ska)}$ and similar solutions for transverse displacements in y and z directions of same amplitude $y_0$, but ensuring that $v_0^2 + w_0^2$ is constant gives us a phonon dispersion branch oscillating along the transverse direction with a frequency governed by $\omega_T^0$ propagating along the crystal length direction and rotating its direction in the y-z plane. The transverse vibration spans the area of a circle of radius $y_0$, The radius will also follow the wavelength determined by k along the longitudinal direction. An instantaneous snapshot so visualized is shown in Fig. 3.

These solutions are possible because of the approximate equivalence of complicated quartic equation solutions to harmonic function.

Therefore a knowledge of the longitudinal force constant enables us to express an equivalent force constant for transverse modes.

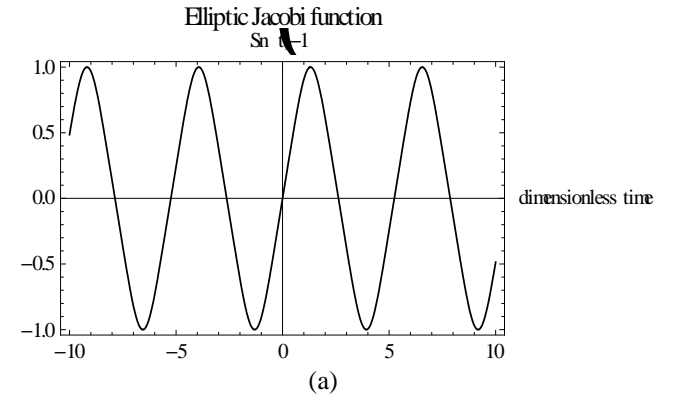

(a)

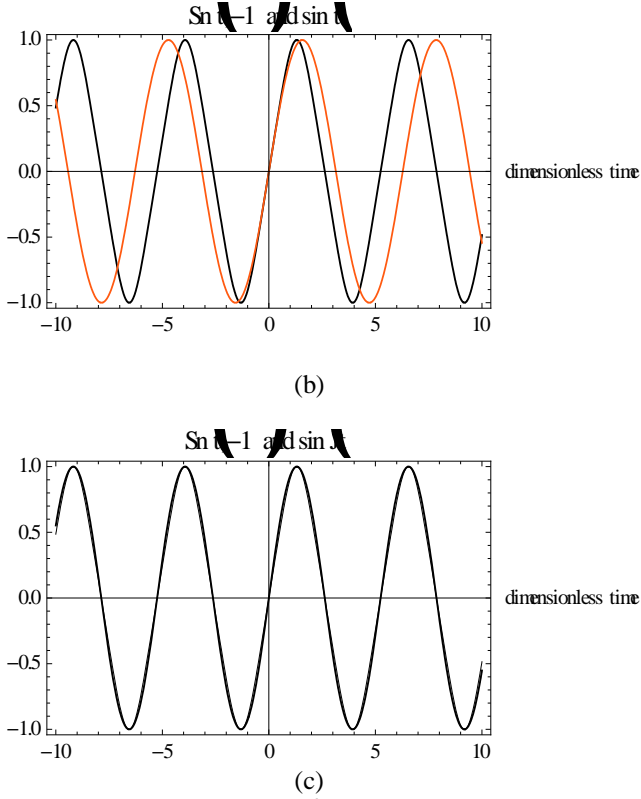

Fig.2. $Y = y/y_0$ vs $\tau = t/\frac{\omega_l^0}{2a}y_0$ as $sn(\tau,-1)$(Black) (a), and $sn(\tau,-1)$ and $\sin(\tau)$(red) in (b), and $sn(\tau,-1)$ and modified $\tau$ as $\sin(J\tau)$ (c).

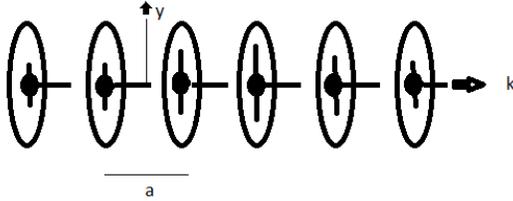

Fig. 3. Propagation of longitudinal as well as transverse waves. The transverse waves allow atoms to vibrate along y directions at the same time while vibrating along $k$ direction.

Eqs.(7) give us the conventional longitudinal phonon dispersion result as

$$\omega^2(kl) = 4\omega^2{}_l{}^0 \sin^2(ka/2), \quad (9)$$

and

$$\omega^2(kT) = 4\omega^2{}_T{}^0 \sin^2(ka/2), \quad (10)$$

Here $l$ stands for longitudinal and T for transverse.

Both transverse branches are degenerate, however they are transverse in polarization to each other and thus the phonon wave propagates with a longitudinal frequency, superimposed with a circularly rotating, transversely vibrating with amplitude $y_0$, a low frequency mode. The transverse frequency is amplitude dependent and thus temperature dependent. The velocity of transverse phonons is very small in comparison to the longitudinal propagation at all $k$ values. The wave vector $k$ is the same as that defined for longitudinal motion.

Further because of the harmonic character of all modes, identified by a polarization $j$, the quantized modes also get similar meaning, with energy written as

$$E = \sum_{k,j}[n_{kj} + 1/2]\hbar\omega(kj), \quad (11)$$
$k = 1, N$ and $j$ includes $l$ and $T$.

Helmholtz free energy can be defined similarly. However the results of temperature dependence of various $\omega$ even in harmonic approximation will contribute even in the absence of explicit anharmonicity as a small component from quartic energy term, $\frac{1}{2}K_T y^4 \sim k_B T$ will result in temperature dependence of $y_0 \sim T^{1/4}$. That will reflect in modification of those thermodynamical properties which involve temperature derivatives of the free energy e.g. heat capacity and entropy.

Similarly, thermal expansion involves Gruneisen parameters which require

$$\gamma = -\frac{\partial \ln(\omega)}{\partial \ln(V)} = -\frac{V}{\omega}\frac{\partial \omega}{\partial V} \quad (12)$$

And for $1-d$ systems, this reduces to
$\gamma = -\frac{L}{\omega}\frac{\partial \omega}{\partial L}$, where $L$ is appropriate length. Thus for transverse modes,

$$\gamma_T = -\frac{y}{\omega_T}\frac{\partial \omega_T}{\partial y} = -1 \quad (13)$$

For longitudinal mode contribution,

$$\gamma_l = -\frac{a}{\omega}\frac{\partial \omega}{\partial a} \sim -\frac{1}{2}a\frac{\frac{\partial^3 \varphi}{\partial a^3}}{\frac{\partial^2 \varphi}{\partial a^2}}. \quad (14)$$

$\frac{\partial^3 \varphi}{\partial a^3}$ which is cubic anharmonic force constant and is usually negative, making $\gamma_l$ a positive quantity. Eq. (14) is conventional Gruneisen parameter originating from anharmonic potential and is absent (so is thermal expansion) in a purely harmonic crystal.

For $3-d$ crystals, the thermal expansion is easily obtained from minimization of the static lattice energy and the Free energy and is given by [2]

$$\in = \frac{1}{2VB}\sum_{kj}\gamma_j \hbar\omega(kj)\coth\left(\frac{\hbar\omega(kj)}{2k_B T}\right) \quad (15)$$

Here B is the bulk modulus and V is the volume of a unit cell. In case of 1-d, we can replace $VB = a^2 N K_l$ which also becomes $aYN$ where Y is Young's modulus.

With the additional contribution of a negative value of $\gamma_T$, the positive contribution from $\gamma_l$ dominating at high temperatures when $\omega_l$ dominates, a competition between the two modes will be responsible for a net thermal expansion contribution. We will see a negative thermal expansion from all such low dimensional systems where the transverse mode is dependent on longitudinal stress at low temperatures. This will also be the case for graphene[3-8]. Sevik [7] discusses in detail various other first principle calculations as well as details some measurements for graphene clearly showing significant negative thermal expansion for graphene and not for graphite. We now see that because of Eq. (13), even in case of purely harmonic low dimensional crystal, there will be a small negative thermal expansion.

A detailed numerical estimate which is quite convenient for one dimensional phonon dispersions can be carried out over longitudinal and transverse modes to realize this. The transverse frequencies remain systematically low by a factor $1.19814 \frac{y_0}{2a}$ as compared to longitudinal modes. The effect of this makes significantly large contribution at low temperatures. For example, the phonon occupancy is highly populated by transverse phonons and low temperature heat capacity will be dominated by contribution from transverse phonons. In general, transverse acoustic modes are going to be playing significant role, especially in those conditions when they are not arising from the bulk, but from the longitudinal interactions.

The present work demonstrates that conventional models of one and two dimensional materials end up in oversimplification especially when applied to interpret the phonon related problems. A simple correction to deal with such systems more realistically is possible. The transverse motions which are derivable from longitudinal motions is the key to this correction. More interestingly, such incorporation results in amplitude dependent frequencies in the transverse directions giving rise to an additional Gruneisen parameter with a value of -1. Since the amplitude will be temperature dependent, these transverse harmonic phonons are weak but show temperature dependence. This analysis is possible because we have been able to reformulate the quartic nonlinear equations which occur essentially, into a quadratic equivalence, though approximately. The quartic nonlinear equations have exact solutions too in Jacobi elliptic functions. Their equivalence to quadratic is achieved approximately by a scale factor identified as a constant $J = 1.19814$. It is further speculated that when cubic anharmonicity is considered, the quartic term will combine and reduce the effect of cubic anharmonicity, leading to interesting softening in phonon phonon interaction resulting in longer collision times and thus raising thermal conductivity. Although *abinitio* calculations always have the possibility to provide these results, but the conceptual part is not evident. It is possible to appreciate why one dimension and two dimension systems have this peculiar behavior. The transverse motion in these cases results from only longitudinal or surface motions. As soon as an additional chain or surface as multilayer graphene is available, such effects will be lost as inter layer interaction provides the features which are like bulk. This motivates a new look at many of the thermodynamical properties of low dimensional systems. Details of such calculations will be the focus of future publication.

VKJ is superannuated faculty and appreciates the Department of Physics at Panjab University to provide facilities and platform. He also gratefully appreciates Professor K N Pathak who provided direction to use Mathematica.

[1] https://wwwx.cs.unc.edu/~snape/publications/mmath/dissertation.pdf by J Snape, retrieved Oct 4, 2015.

[2] V. K. Jindal and J. Kalus, *Phys. Status Solidi* **133**, 89–99 (1986).

[3] Sarita Mann, Pooja Rani, Ranjan Kumar and V.K.Jindal, (60[th] DAE-SSPS 2015, to appear as AIP proceedings.)

[4] Yoon D, Son Y-W and Cheong H, *Nano Lett* **11**, 3227 (2011).

[5] N. Mounet, and N. Marzari, *Phys. Rev.* B **71**, 205214 (2005).

[6] Jiang J W, Wang J S and Li B, *Phys. Rev.* B **80** 205429 (2009).

[7] C. Sevik, *Phys. Rev.* B **89**, 035422 (2014).

[8] Wenzhong Bao, Feng Miao, Zhen Chen, Hang Zhang, Wanyoung Jang, Chris Dames and Chun Ning Lau, *Nature Nanotechnolog* 4, 562 (2009)